\documentclass[10pt,twocolumn,aps,prd,preprintnumbers,showpacs,superscriptaddress,nofootinbib,amsmath,amssymb,floats,floatfix,showkeys,notitlepage,longbibliography]{revtex4-1}
\usepackage[T1]{fontenc}
\usepackage[utf8]{inputenc}
\usepackage{newtxtext}
\usepackage{newtxmath}

\usepackage{amsmath,amssymb,amsfonts}

\usepackage{xcolor,graphicx,float}
\usepackage{orcidlink}
\usepackage{mdframed}
\usepackage{hyperref} 

\newcommand{\gc}{g_{\!c}}

\begin{document}

\title{Acceleration Radiation of Freely Falling Atoms: Nonlinear Electrodynamic Effects}
\author{Ali \"Ovg\"un \orcidlink{0000-0002-9889-342X}}
\email{ali.ovgun@emu.edu.tr}
\affiliation{Physics Department, Faculty of Arts and Sciences, Eastern Mediterranean University, Famagusta, 99628 North Cyprus via Mersin 10, Turkiye.}

\author{Reggie C. Pantig \orcidlink{0000-0002-3101-8591}} 
\email{rcpantig@mapua.edu.ph}
\affiliation{Physics Department, School of Foundational Studies and Education, Map\'ua University, 658 Muralla St., Intramuros, Manila 1002, Philippines.}

\author{Bobomurat Ahmedov \orcidlink{0000-0002-1232-610X}} 
\email{ahmedov@astrin.uz}
\affiliation{Institute for Advanced Studies, New Uzbekistan University,
Movarounnahr str. 1, Tashkent 100000, Uzbekistan
.}
\affiliation{Institute of Theoretical Physics, National University of Uzbekistan, Tashkent 100174, Uzbekistan.}
\affiliation{School of Physics, Harbin Institute of Technology, Harbin 150001, People’s Republic of China.}
\author{Uktamjon Uktamov\orcidlink{0009-0003-0423-2474}} 
\email[Corresponding Author:]{uktam.uktamov11@gmail.com}
\affiliation{School of Physics, Harbin Institute of Technology, Harbin 150001, People’s Republic of China.}
\affiliation{Tashkent University of Applied Sciences, Gavhar Str. 1, Tashkent 100149, Uzbekistan.}
\affiliation{Tashkent State Technical University, Tashkent 100095, Uzbekistan}
\affiliation{Institute for Advanced Studies, New Uzbekistan University,
Movarounnahr str. 1, Tashkent 100000, Uzbekistan
.}

\date{\today}

\begin{abstract}
Motivated by the work of Scully \textit{et al.} [ \textcolor{blue}{Proc. Nat. Acad. Sci. 115, 8131 (2018)}] and Camblong \textit{et al.}[ \textcolor{blue}{Phys. Rev. D 102, 085010 (2020)}], we investigate horizon-brightened acceleration radiation (HBAR) for freely falling two-level atoms in the geometry of a Bardeen regular black hole. Building on the quantum-optics approach to acceleration radiation and its near-horizon conformal quantum mechanics (CQM) structure, we show that the dominant physics is again governed by an inverse-square potential in the radial Klein-Gordon equation, with an effective coupling fixed by the Bardeen surface gravity. Using geodesic expansions and a near-horizon CQM reduction of the scalar field, we derive the excitation probability for atoms falling through a Boulware-like vacuum in the presence of a stretched-horizon mirror. The resulting spectrum is Planckian in the mode frequency, with a temperature determined by the Bardeen Hawking temperature. We analyze how the regular core parameter controls the strength of the radiation and demonstrate that the excitation probability is strongly suppressed as the geometry approaches the extremal (cold remnant) limit. Numerical results illustrate the dependence of the spectrum on the Bardeen parameter and on the atomic transition frequency.
\end{abstract}

\maketitle

\section{Introduction}

The deep interplay between gravitation, quantum field theory, and thermodynamics was first revealed in the seminal discovery that black holes radiate thermally with a temperature proportional to their surface gravity \cite{ GibbonsHawking,Hawking}. In Hawking's calculation, quantum fields on a fixed black hole background produce a flux of particles at infinity, while in the Unruh effect, uniformly accelerated observers in flat spacetime detect a thermal bath in the Minkowski vacuum \cite{Unruh:1976db,FullingUnruh}. These phenomena are now understood as cornerstone examples of quantum field theory in curved spacetime and non-inertial frames \cite{BirrellDavies,WaldQFT,WaldGR,DeWittQFT}.

A complementary and operational perspective on these effects is provided by particle-detector models, in which localized two-level systems (or Unruh-DeWitt detectors) couple to quantum fields along prescribed worldlines. Such detectors have been used to probe Unruh and Hawking thermality in a variety of settings, including uniformly accelerated trajectories, cosmological spacetimes, and black hole geometries, with an emphasis on response functions, entanglement harvesting, and the weak equivalence principle \cite{Tjoa:2018xre,Lopp:2018lxl,You:2018qgr,Fulling:2018lez,Chakraborty:2019ltu,MasoodASBukhari:2023flr}. 

Within this framework, a particularly fruitful line of research is the quantum-optics program initiated by Scully and collaborators, who recast Hawking-Unruh physics in terms of atoms interacting with quantum fields in curved backgrounds and accelerated frames \cite{Scully:2017utk,Svidzinsky:2018jkp,Svidzinsky:2019jqr}. A central result of this program is the notion of \emph{horizon-brightened acceleration radiation} (HBAR): when a two-level atom falls freely into a Schwarzschild black hole through a Boulware-like vacuum in the presence of a near-horizon mirror, it can still become excited while emitting radiation with a Planckian spectrum characterized by the Hawking temperature, despite the absence of outgoing Hawking flux at future null infinity \cite{Scully:2017utk,Camblong:2020pme}. This effect arises from the relative acceleration between the atom and the field modes constrained by the mirror, and can be traced to a universal near-horizon conformal quantum mechanics (CQM) structure in the radial Klein-Gordon equation \cite{Camblong:2020pme,Azizi:2020gff,Azizi:2021qcu,Azizi:2021yto,Das:2022qcr,Das:2022qpx,Good:2021ffo,Ordonez:2025sqp}.

The near-horizon CQM perspective shows that, for a wide class of static and stationary black holes, the dominant physics of s-wave modes is governed by an inverse-square potential whose effective coupling is fixed by the surface gravity. This insight has led to a series of extensions of the HBAR setup: to rotating and charged black holes \cite{Azizi:2020gff,Sen:2023zfq}, to general classes of static spherically symmetric metrics beyond Schwarzschild \cite{Sen:2022cdx}, to braneworld black holes \cite{Das:2023rwg}, and to backgrounds with external fields, such as dark energy or dark matter distributions \cite{Bukhari:2022wyx,Bukhari:2023yuy}. More recent work has pushed the formalism to derivative-coupled detectors and modified gravity scenarios \cite{Pantig:2025okn,Das:2025rzz,Rahaman:2025grm,Rahaman:2025mrr,Tang:2025eew}, to Lorentz-violating spacetimes, and  resonantly driven setups where redshift-enhanced and Floquet-type acceleration radiation can emerge \cite{Pantig:2025igg,Masood:2024glj,Wang:2025jvb,Pan:2024jix}.

A parallel development has focused on the thermodynamic and entropic aspects of acceleration radiation. Building on master-equation techniques and the CQM structure, Azizi \textit{et al.} have shown how the coarse-grained dynamics of a single cavity mode interacting with a flux of infalling atoms leads to entropy production and area-law-like relations that mirror Hawking's original thermodynamic picture \cite{Azizi:2021qcu,Azizi:2021yto}. Subsequent works have explored HBAR entropy and its quantum-gravity-motivated corrections in quantum-corrected and renormalization-group improved black hole backgrounds \cite{Sen:2022tru,Jana:2024fhx,Jana:2025hfl,Ovgun:2025isv,Eissa:2025vhs}, including inverse-logarithmic and generalized-uncertainty-principle (GUP) corrections, as well as causal-diamond geometries. These studies strengthen the view that the HBAR setup provides a versatile quantum-optical laboratory for probing horizon thermality, the equivalence principle, and possible quantum-gravity imprints \cite{Chatterjee:2021fue,Chatterjee:2021fsg,Barman:2021oum,Barman:2021kwg,Liu:2019soq,Barman:2023rhd,Allen:2019muo}.

At the same time, there is growing interest in connecting these theoretical developments to experiments and analogue systems. Proposals range from Ramsey interferometry and cavity-QED schemes designed to witness acceleration radiation \cite{Costa:2020asn,Lopes:2021ajs,Gregori:2023tun}, to studies of dynamical Casimir and gravitational-wave-induced effects in quantum optical settings \cite{Sorge:2018zfd,Sorge:2023tyn,Chatterjee:2022pkj}, to analogue-gravity platforms based on water waves and optical media that mimic horizon kinematics and surface gravity \cite{BarceloVisserLiberati,Rozenman:2024ymh,Xu:2021cbw}. Together with more formal advances on nonlocal fields, nonvacuum states, and Wigner distributions in Rindler spacetime \cite{Prokhorov:2019sss,Das:2022qcr,Das:2022qpx,Dubey:2024abz,Stargen:2025siq}, these efforts suggest that acceleration radiation and HBAR may provide experimentally relevant windows into quantum aspects of spacetime.

However, most HBAR studies to date have focused on singular black holes or on effective quantum-corrected metrics that still retain a central curvature singularity, albeit softened or surrounded by a new structure \cite{Camblong:2020pme,Sen:2022tru,Sen:2022cdx,Jana:2024fhx,Ovgun:2025isv}. Lorentzian–Euclidean black holes provide a nonsingular framework for avoiding singularities, where atemporality (rooted in conservation laws) crucially influences the causal structure, null-geodesic dynamics, and matter accretion \cite{Capozziello:2024ucm,DeBianchi:2025bgn,Capozziello:2025wwl}. In contrast, \emph{regular} black holes--and in particular the Bardeen solution, which arises as an effective geometry of Einstein gravity coupled to nonlinear electrodynamics—provide de~Sitter-like cores with no curvature singularity at the origin \cite{Bambi:2013ufa,Ayon-Beato:2000mjt}. The Bardeen parameter $g$ controls the size of the regular core and interpolates between a Schwarzschild-like regime ($g\to 0$) and an extremal zero-temperature remnant in which the outer and inner horizons coincide. Regular black holes are widely used as phenomenological models of short-distance quantum-gravity effects and as candidate endpoints of Hawking evaporation. However, their imprint on acceleration radiation, detector response, and HBAR entropy has not been systematically explored.

In this work, we bring the HBAR and near-horizon CQM program to the realm of Bardeen regular black holes. We consider a neutral scalar field in the Bardeen geometry and a two-level atom freely falling through a Boulware-like vacuum in the presence of a stretched-horizon mirror, following the quantum-optics formulation of Refs.~\cite{Scully:2017utk,Camblong:2020pme,Azizi:2021qcu,Azizi:2021yto,Ordonez:2025sqp}. Our goals are twofold. First, we show that the near-horizon CQM structure responsible for HBAR thermality persists in the Bardeen spacetime, with an effective inverse-square potential whose coupling is fixed by the Bardeen surface gravity and hence by the core parameter $g$. This leads to a Planckian excitation spectrum for the falling atom characterized by the Bardeen Hawking temperature and a strong suppression of the radiation as the extremal, cold-remnant limit is approached. Second, we construct a coarse-grained master equation and an associated HBAR entropy for a selected cavity mode in this regular background, and investigate how singularity resolution and the de~Sitter-like core deform the Wien-type relations and area-law behavior found in earlier HBAR entropy analyses \cite{Azizi:2021qcu,Azizi:2021yto,Sen:2022tru,Jana:2024fhx,Jana:2025hfl,Ovgun:2025isv,Eissa:2025vhs}. In this way, the Bardeen parameter acts as a tunable ``regularization scale'' controlling both the strength of acceleration radiation and the structure of its entropy, providing a quantum-optical probe of regular black hole remnants and their thermodynamic properties.

The paper is organized as follows. In Sec.~\ref{sec:Bardeen} we review the Bardeen metric, its horizon structure, and the associated surface gravity and Hawking temperature, emphasizing the role of the parameter $g$ and the extremal limit. In Sec.~\ref{sec:geo} we discuss the geodesic motion of freely falling atoms, derive the near-horizon expansions of proper and coordinate time, and identify the leading behavior relevant for the detector response. In Sec.~\ref{sec:CQM} we derive the near-horizon CQM equation for the scalar field in the Bardeen spacetime and obtain the outgoing mode structure. Section~\ref{sec:spectrum} is devoted to the calculation of the excitation probability using time-dependent perturbation theory, leading to a Planckian spectrum with the Bardeen temperature. We then present and interpret numerical results for the dependence of the excitation probability on the Bardeen parameter and on the atomic frequency.  In Sec.~\ref{sec:HBARentropy} we construct a coarse-grained master equation for a
single cavity mode, derive the associated HBAR entropy in the Bardeen background,
and discuss how the regular core modifies the energy-entropy flux and its relation
to the horizon area. In Sec.~\ref{conclusion} we conclude.

We work in natural units with $c=\hbar=k_B=G=1$ unless otherwise stated. The Bardeen parameter of the geometry is denoted by $g$ (core scale), whereas the atom-field coupling is denoted by $\gc$.

\section{Bardeen black hole spacetime}
\label{sec:Bardeen}

We consider a static, spherically symmetric, regular (nonsingular) black hole of Bardeen type. Its line element can be written in standard Schwarzschild-like coordinates as \cite{Bambi:2013ufa,Ayon-Beato:2000mjt}
\begin{align}
    ds^{2}
    = - F(r)\,dt^{2}
      + \frac{dr^{2}}{F(r)}
      + r^{2} d\theta^{2}
      + r^{2} \sin^{2}\theta\, d\phi^{2},
    \label{eq:metric}
\end{align}
with metric function
\begin{equation}
    F(r) = 1 - \frac{2 M(r)}{r},
    \qquad
    M(r) = m\,\frac{r^{3}}{\bigl(r^{2} + g^{2}\bigr)^{3/2}},
    \label{eq:F_Bardeen}
\end{equation}
where $m$ is the ADM mass and $g$ is the Bardeen parameter (core scale).

The mass function $M(r)$ interpolates between an effective de Sitter core at small radii and the usual Schwarzschild asymptotics at large radii. For $r\ll g$, one finds $M(r)\sim m\,r^{3}/g^{3}$, so that $F(r)\approx 1 - (2m/g^{3})\,r^{2}$, corresponding to a regular de Sitter-like behavior at the origin. For $r\gg g$, the mass function tends to $m$ and the geometry approaches the Schwarzschild solution. Thus, $g$ controls the departure from the classical singular black hole and can be viewed as a phenomenological parameter encoding short-distance regularization.

The event horizon is located at $r=r_g$, defined implicitly by
\begin{equation}
    F(r_g) = 0
    \quad\Longrightarrow\quad
    m = \frac{\bigl(r_g^{2} + g^{2}\bigr)^{3/2}}{2 r_g^{2}}.
    \label{eq:rg_def}
\end{equation}
For given $(m,g)$ this equation determines the horizon radius; conversely, it is often convenient to trade $m$ for $r_g$ via Eq.~\eqref{eq:rg_def}, which will be used extensively in our near-horizon analysis.

To characterize the near-horizon behavior of the metric function, we compute its first and second derivatives:
\begin{align}
    F'(r)
    &= \frac{2 m r \bigl(r^{2} - 2 g^{2}\bigr)}{\bigl(r^{2} + g^{2}\bigr)^{5/2}},
    \label{eq:Fprime_general}\\
    F''(r)
    &= \frac{30 m g^{2} r^{2}}{\bigl(r^{2} + g^{2}\bigr)^{7/2}}
       - \frac{4 m}{\bigl(r^{2} + g^{2}\bigr)^{3/2}}.
    \label{eq:Fsecond_general}
\end{align}
Evaluating at the horizon $r=r_g$ and using Eq.~\eqref{eq:rg_def}, we obtain compact expressions
\begin{eqnarray}
    F'_g \equiv F'(r_g)
    = \frac{2 m r_g (r_g^{2} - 2 g^{2})}{(r_g^{2}+g^{2})^{5/2}}
    = \frac{r_g^{2} - 2 g^{2}}{r_g \bigl(r_g^{2} + g^{2}\bigr)},
    \qquad \\
    F''_g \equiv F''(r_g)
    = \frac{15 g^{2}}{\bigl(r_g^{2} + g^{2}\bigr)^{2}}
      - \frac{2}{r_g^{2}}.
    \label{eq:Fg}
\end{eqnarray}
The sign and magnitude of $F'_g$ control the causal nature of the horizon. For $r_g^{2}>2g^{2}$ one has $F'_g>0$ and the horizon is nonextremal, while $r_g^{2}=2g^{2}$ corresponds to an extremal configuration with $F'_g=0$, where the outer and inner horizons coincide.

The surface gravity and Hawking temperature of the Bardeen black hole are therefore
\begin{equation}
    \kappa = \frac{1}{2} F'_g,
    \qquad
    T_{H}^{(\mathrm{B})}
    = \frac{\kappa}{2\pi}
    = \frac{\tfrac{1}{2} r_g^{2} - g^{2}}{2\pi\,r_g\bigl(r_g^{2} + g^{2}\bigr)}.
    \label{eq:kappa_T}
\end{equation}
The limit $g\to 0$ recovers the Schwarzschild surface gravity and Hawking temperature, while increasing $g$ progressively lowers $T_{H}^{(\mathrm{B})}$. For $r_g^{2} > 2 g^{2}$ the horizon is nonextremal ($\kappa>0$), whereas at
$r_g^{2} = 2 g^{2}$ the black hole becomes extremal with $\kappa=0$ and
$T_{H}^{(\mathrm{B})}=0$, providing a simple model of a cold remnant. One of our main goals below will be to track how this change in surface gravity is encoded in the acceleration radiation detected by freely falling atoms.

\begin{figure}
    \centering
\includegraphics[width=1\linewidth]{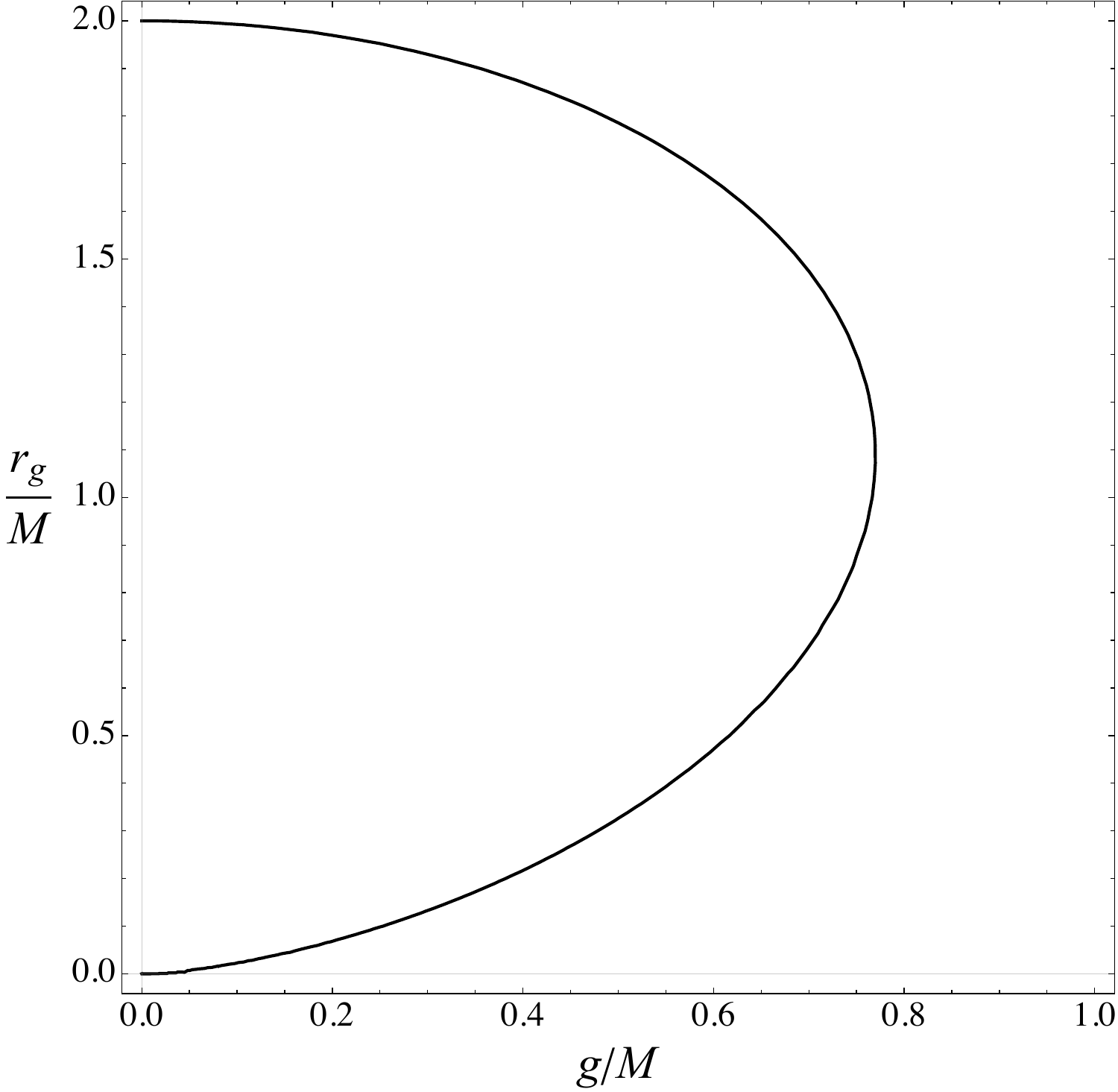}\hfill
    \caption{Dependence of the horizon radius $r_g$ on Bardeen parameter $g$.}
    \label{fig:rg}
\end{figure}

\section{Geodesic equations}
\label{sec:geo}

The spacetime trajectories of freely falling atoms are governed by the geodesic equation
\begin{equation}
    \frac{d^{2}x^{\mu}}{d\tau^{2}}
    + \Gamma^{\mu}_{\rho\sigma}
      \frac{dx^{\rho}}{d\tau}\frac{dx^{\sigma}}{d\tau}
    = 0,
    \label{eq:geo}
\end{equation}
where $\tau$ is the proper time and $\Gamma^{\mu}_{\rho\sigma}$ are the Christoffel symbols associated with the metric~\eqref{eq:metric}. We are interested in radial infall, appropriate for atoms dropped from rest far from the black hole.

For the static and spherically symmetric metric~(\ref{eq:metric}), spherical symmetry allows us to restrict the motion to the equatorial plane $\theta=\pi/2$, and azimuthal symmetry implies that we may consider trajectories with vanishing angular momentum, so that $\dot{\theta} = \dot{\phi} = 0$ and the motion is purely radial. The conserved energy per unit mass associated with the Killing vector $\partial_t$ is
\begin{equation}
    e = F(r)\,\frac{dt}{d\tau}.
    \label{eq:eng}
\end{equation}
The normalization $u^{\mu}u_{\mu}=-1$ of the four-velocity $u^{\mu} = dx^{\mu}/d\tau$ then yields
\begin{equation}
    \left(\frac{dr}{d\tau}\right)^{2}
      = e^{2} - F(r),
    \qquad
    \left(\frac{dr}{dt}\right)^{2}
      = \left(\frac{F(r)}{e}\right)^{2}\bigl[e^{2} - F(r)\bigr].
    \label{eq:con}
\end{equation}
These relations have the familiar interpretation: the radial velocity in proper time is controlled by the ``energy balance'' between $e^{2}$ and the redshift function $F(r)$, while the radial velocity with respect to the coordinate time $t$ is suppressed near the horizon by the vanishing of $F(r)$.

Integrating Eq.~\eqref{eq:con}, we can write the proper time and coordinate time elapsed during the infall from some initial radius $r_i$ to a final radius $r_f$ as
\begin{equation}
    \tau
    = -\int_{r_i}^{r_f} dr
        \frac{1}{\sqrt{e^{2} - F(r)}},
    \,\,\,
    t
    = -\int_{r_i}^{r_f} dr
        \frac{e}{F(r)\,\sqrt{e^{2} - F(r)}},
    \label{eq:time}
\end{equation}
where we have chosen the minus sign to describe infall ($r$ decreasing with increasing $\tau$ and $t$). For an atom released from rest at infinity in an asymptotically flat spacetime, one has $e=1$.

Near the event horizon $\mathcal{H}$, with $r\sim r_g$, it is convenient to introduce the shifted coordinate $x = r - r_g$, which measures the radial distance to the horizon. The near-horizon expansions of $F(r)$ and its derivatives can then be organized as a hierarchical series in powers of $x$ \cite{Tang:2025eew}:
\begin{align}
    F(r)
    &\overset{(\mathcal{H})}{\sim}
      F'_g\,x\left[1 + \mathcal{O}(x)\right], \notag\\
    F'(r)
    &\overset{(\mathcal{H})}{\sim}
      F'_g\left[1 + \mathcal{O}(x)\right], \notag\\
    F''(r)
    &\overset{(\mathcal{H})}{\sim}
      F''_g\left[1 + \mathcal{O}(x)\right],
    \label{eq:taly}
\end{align}
where $F'_g$ and $F''_g$ are given in Eq.~(\ref{eq:Fg}) and the symbol $\overset{(\mathcal{H})}{\sim}$ denotes the near-horizon expansion. The leading term in $F(r)$ is linear in $x$ for a nonextremal horizon; this linear behavior is crucial for the emergence of an effective Rindler structure and the associated conformal quantum mechanics. \paragraph{Domain of validity (nonextremal horizon).}
Throughout the near-horizon analysis we assume a \emph{nonextremal} outer horizon,
\begin{equation}
F'_g \neq 0 \quad \Leftrightarrow \quad r_g^2>2g^2 \, ,
\end{equation}
so that $F(r)\sim F'_g (r-r_g)$ and the local geometry is Rindler-like.
The exactly extremal Bardeen configuration ($r_g^2=2g^2$, $\kappa=0$) has
$F(r)\sim \tfrac12 F''_g (r-r_g)^2$ and requires a separate near-horizon treatment.
Our statements about ``approaching extremality'' should therefore be understood as
the limit $\kappa\to 0^+$ within the nonextremal family.

Using these expansions, one can evaluate Eq.~(\ref{eq:time}) near the horizon. The main source of divergence in $t$ is the factor $1/F(r)$, whose expansion we make explicit:
\begin{equation}
    \frac{1}{F(r)}=\frac{1}{F'_g x}\left(1-\frac{F''_g}{2F'_g}x+\mathcal O(x^2)\right),
    \label{eq:oneoverFexpansion}
\end{equation}
while the square root $\sqrt{e^{2}-F(r)}$ can be expanded as a regular power series in $x$ since $F(r)\to 0$ at the horizon. From these ingredients, keeping terms up to first order in $x$, one finds
\begin{align}
    \tau
    &= -\frac{1}{e} x
       + \mathrm{const.}
       + \mathcal{O}(x^{2}),
    \label{eq:taylp}\\
    t
    &= -\frac{1}{F'_g}\,\ln x
       - C\,x
       + \mathrm{const.}
       + \mathcal{O}(x^{2}),
    \label{eq:taylt}
\end{align}
where the constant $C$ depends on $e$ and on the geometry through the second derivative of $F(r)$:
\begin{equation}
    C = \frac{1}{2}\left[
          \frac{1}{e^{2}} - \frac{F''_g}{\bigl(F'_g\bigr)^{2}}
        \right].
    \label{eq:ccc}
\end{equation}
Equation~\eqref{eq:taylp} shows that the proper time to reach the horizon is finite, while Eq.~\eqref{eq:taylt} displays the familiar logarithmic divergence of the coordinate time $t$, reflecting the infinite redshift as seen by a static observer at infinity. In the immediate vicinity of the horizon, the $\mathcal{O}(x^{2})$ terms can be safely neglected in the calculation of the detector response, which is dominated by the leading behavior of $\tau(x)$ and $t(x)$.

\section{Near-horizon conformal quantum mechanics equation}
\label{sec:CQM}

We now turn to the quantum field that interacts with the falling atom \cite{Azizi:2021qcu,Azizi:2021yto}. We consider a minimally coupled, massless scalar field obeying the Klein-Gordon equation
\begin{align}
    \frac{1}{\sqrt{-g}}\,
    \partial_{\mu}\!\left(
      \sqrt{-g}\,g^{\mu\nu}
      \partial_{\nu}\Phi
    \right)
    = 0,
    \label{eq:kg}
\end{align}
on the Bardeen background~(\ref{eq:metric}). Spherical symmetry allows for a separation into spherical harmonics; however, in the near-horizon region and for our purposes, the dominant contribution to the detector response comes from the s-wave sector. Moreover, the presence of a reflecting mirror close to the horizon renders
greybody factors irrelevant for the excitation process, so an effective
$(1+1)$-dimensional description is appropriate.
For the $s$-wave we define $\psi(t,r)\equiv r\,\Phi(t,r)$, after which the
angular barrier is absent and the radial equation reduces to the $(t,r)$ sector.

Restricting to the $(t,r)$ sector and s-wave modes, the field equation reduces to the effective $(1+1)$-dimensional form,
\begin{equation}
   -\frac{1}{F(r)}\frac{\partial^{2}\Phi}{\partial t^{2}}
   +\frac{\partial}{\partial r}\!\left(F(r)\,\frac{\partial\Phi}{\partial r}\right)
   = 0.
   \label{eq:app1_B}
\end{equation}
This equation already exhibits the key role of the redshift function $F(r)$: time derivatives are weighted by $1/F(r)$, which diverges at the horizon, while radial derivatives are multiplied by $F(r)$, which vanishes there.

We expand the field in modes as \cite{Camblong:2020pme}
\begin{equation}
    \Phi(t,r)
    = \sum \bigl[\hat{a}\,\phi(r,t) + \mathrm{H.c.}\bigr],
    \label{eq:app2}
\end{equation}
with separable solutions
\begin{equation}
    \phi(t,r) = \chi(r)\,u(r)\,e^{-i\nu t},
    \label{eq:svs}
\end{equation}
where $\nu$ is the mode frequency. The function $\chi(r)$ is chosen to remove the first-derivative term in the radial equation and thus cast it into a Schr\"odinger-like form. The standard choice,
\begin{equation}
    \chi(r)
    = \exp\!\left(
         -\frac{1}{2} \int \frac{F'(r)}{F(r)}\,dr
      \right)
    = \bigl[F(r)\bigr]^{-1/2},
    \label{eq:app4}
\end{equation}
ensures that the radial part of the Klein-Gordon equation is cast into
\begin{equation}
    u''(r) + V_{(D)}(r;\nu)\,u(r) = 0,
    \label{eq:skg}
\end{equation}
with an effective potential $V_{(D)}(r;\nu)$ that incorporates both the frequency dependence and the geometric information encoded in $F(r)$ and its derivatives. The explicit form of $V_{(D)}$ is used to extract the leading near-horizon behavior.

Near the horizon $\mathcal{H}$, using the expansions~(\ref{eq:taly}), the radial equation simplifies drastically. The leading contributions in the potential behave as an inverse square in the proper distance to the horizon, and the radial equation takes the CQM form \cite{Camblong:2003mz}
\begin{equation}
    u''(x)
    + \frac{\lambda_{\mathrm{eff}}}{x^{2}}
      \bigl[1+\mathcal{O}(x)\bigr]\,u(x)
    = 0,
    \label{eq:cqm}
\end{equation}
where $x=r-r_g$ and the effective coupling is
\begin{equation}
    \lambda_{\mathrm{eff}} = \frac{1}{4} + \Theta^{2},
    \qquad
    \Theta = \frac{\nu}{F'_g}.
    \label{eq:lambda_Theta}
\end{equation}
Equation~(\ref{eq:cqm}) shows that the dominant near-horizon physics is governed by the long-range conformal quantum mechanics (CQM) potential
\begin{equation}
    V_{\mathrm{(D)}}(x)
    = \frac{\lambda_{\mathrm{eff}}}{x^{2}}.
\end{equation}
The CQM structure is a universal consequence of the linear behavior of $F(r)$ near a nonextremal horizon and underlies the emergence of a thermal factor in the detector response. In the present context, the Bardeen parameter $g$ enters only through $F'_g$ and thus through $\Theta$ and $\lambda_{\mathrm{eff}}$, providing a direct link between the regular core scale and the effective CQM coupling.

The outgoing, horizon-regular solution of Eq.~\eqref{eq:cqm} is
\begin{equation}
    u(x)
    = x^{\frac{1}{2}+\sqrt{\frac{1}{4}-\lambda_{\mathrm{eff}}}}
    = \sqrt{x}\,x^{i\Theta},
    \label{eq:ux}
\end{equation}
where in the last step we used $\lambda_{\mathrm{eff}} = 1/4+\Theta^{2}$, so that the exponent is purely imaginary. This structure is characteristic of scattering in an inverse-square potential and reflects the presence of an underlying $SL(2,\mathbb{R})$ symmetry in the near-horizon dynamics.

Equivalently, the near-horizon mode can be written in an Eddington-Finkelstein-like form as
\begin{equation}
    \phi(t,r)\simeq \frac{1}{\sqrt{F'_g}}\,e^{-i\nu\,(t-r_\ast)},
    \qquad r_\ast=\frac{1}{F'_g}\ln x + \mathcal O(x),
    \label{eq:outgoingmode}
\end{equation}
where $r_\ast$ is the tortoise coordinate. The logarithmic behavior of $r_\ast$ near the horizon is again a manifestation of the Rindler-like structure of the metric in that region. The overall normalization $1/\sqrt{F'_g}$ will enter only as a prefactor in the excitation probability and does not affect the thermal nature of the spectrum.

\section{Particle spectrum of acceleration radiation}
\label{sec:spectrum}

We now couple the scalar field to a two-level atom following a geodesic trajectory and compute the probability for the atom to become excited while emitting a photon. For clarity, we recall the setup schematically: a black hole is located at the center, and a (stretched) mirror is placed just outside the event horizon to shield infalling atoms from Hawking radiation. The mirror enforces a boundary condition on the field that eliminates any outgoing Hawking flux at future null infinity and ensures that the relevant radiation is purely due to the atom-field interaction in the curved background. A two-level dipole atom, acting as a detector, falls freely towards the black hole described by metric~(\ref{eq:metric}). The relative acceleration between the atom and the field modes (which ``see'' the mirror as a fixed boundary) gives rise to acceleration radiation, which can be detected by an observer at asymptotic infinity.

Since no Hawking flux reaches the asymptotic observer due to the mirror, the field is taken to be in a Boulware-like vacuum \cite{Boulware:1974dm},
\begin{equation}
    \hat{a}\,|0_{B}\rangle = 0,
    \label{eq:bwc}
\end{equation}
free of outgoing flux at $\mathcal I^+$. Because the relevant dynamics is local to the near-horizon region with a reflecting boundary, greybody factors associated with the potential barrier at larger radii are irrelevant for the excitation process.

Using time-dependent perturbation theory, the excitation probability for the atom to make a transition from the ground state $|b\rangle$ to the excited state $|a\rangle$ while emitting a field quantum of Killing frequency $\nu$ is
\begin{equation}
    P_{\rm exc}
    = \left|
        \int d\tau\,
        \langle 1_{\nu},a |
          V_{I}(\tau)
        | 0_{B},b\rangle
      \right|^{2},
    \label{eq:p}
\end{equation}
where $\tau$ is the proper time along the atomic trajectory and $|1_{\nu}\rangle$ denotes the one-photon mode with frequency~$\nu$. In natural units, the interaction Hamiltonian in the interaction picture can be written as \cite{ScullyHBAR}
\begin{eqnarray}\label{eq:v}
    V_{I}(\tau)
    &=&\gc
      \bigl[
        \hat{a}_{\nu}\,\phi(r(\tau),t(\tau))
        + \hat{a}^{\dagger}_{\nu}\,\phi^{*}(r(\tau),t(\tau))
      \bigr]\\\nonumber
      &\times&\bigl(
        \sigma_{-} e^{-i\omega\tau}
        + \mathrm{H.c.}
      \bigr),
\end{eqnarray}

where $\sigma_{-}$ is the atomic lowering operator, $\omega$ is the atomic transition frequency, and $\gc$ is the atom-field coupling constant. The structure of Eq.~\eqref{eq:v} is standard in quantum optics: the atom couples to the local field operator evaluated along its worldline, and the $\sigma_{-}$, $\sigma_{+}$ terms encode de-excitation and excitation processes.

Using Eq.~(\ref{eq:p}), only the term proportional to $\hat{a}^{\dagger}_{\nu}\sigma_{+}$ contributes to the excitation probability, because we start from the combined state $|0_{B},b\rangle$ and end in $|1_{\nu},a\rangle$. Therefore, we can write
\begin{equation}
    P_{\rm exc}
    = \gc^{2}
      \Bigl|
        \int d\tau\,
        \phi^{*}(r(\tau),t(\tau))\,e^{i\omega\tau}
      \Bigr|^{2}.
    \label{eq:p1}
\end{equation}
This expression makes it clear that the excitation probability is determined by the overlap between the phase evolution of the field mode along the atomic trajectory and the internal phase evolution of the atom.

The relevant field modes entering Eq.~(\ref{eq:p1}) follow from the CQM equation~(\ref{eq:cqm}). We select the outgoing solution~(\ref{eq:ux}) near the horizon, and use $\chi(r)=[F(r)]^{-1/2}$. Near $\mathcal{H}$ we have
\begin{equation}
    \chi(r)
    = [F(r)]^{-1/2}
    \overset{(\mathcal{H})}{\sim}
      \frac{1}{\sqrt{F'_g x}}\,
      \bigl[1 + \mathcal{O}(x)\bigr],
    \label{eq:xr}
\end{equation}
so that an equivalent near-horizon form of the full mode function is
\begin{equation}
    \phi(r,t)
    \overset{(\mathcal{H})}{\sim}
      \frac{1}{\sqrt{F'_g}}\,e^{-i\nu\,(t-r_\ast)}
    = \frac{1}{\sqrt{F'_g}}\,x^{i\Theta} e^{-i\nu t},
    \label{eq:bohs}
\end{equation}
where $r_\ast=(1/F'_g)\ln x +\cdots$. The free-fall trajectory is given near the horizon by Eqs.~(\ref{eq:taylp})-(\ref{eq:taylt}). Combining these with Eq.~(\ref{eq:bohs}), and keeping only the leading contributions in $x$, one finds
\begin{equation}
    \phi^{*}(r(\tau),t(\tau))\,e^{i\omega\tau}
    \propto
    x^{-i\sigma}\,e^{-isx},
\end{equation}
with
\begin{align}
    \sigma &= 2\Theta
      = \frac{2\nu}{F'_g},
      \label{eq:sigma_B}\\
    s &= C\nu + \frac{\omega}{e}.
      \label{eq:sss_B}
\end{align}
The parameter $\sigma$ encodes the near-horizon redshift of the mode frequency, while $s$ incorporates both the geometric correction $C\nu$ and the contribution from the atomic frequency $\omega$.

Using $d\tau \simeq -dx/e$ near the horizon, we obtain from Eq.~(\ref{eq:p1})
\begin{equation}
    P_{\rm exc}
    = \frac{\gc^{2}}{e^{2}}\,
      \left|
        \int_{0}^{x_{f}} dx\,
        x^{-i\sigma}\,e^{-isx}
      \right|^{2},
    \label{eq:p2}
\end{equation}
where $x_{f}$ denotes the upper boundary of the region where the near-horizon approximation is valid. The integral in Eq.~\eqref{eq:p2} is dominated by the region $x\lesssim 1/s$; for sufficiently large $s$ the main contribution comes from the near-horizon zone, justifying the use of the CQM-based mode structure.

For $s\gg\sigma$ (e.g.\ in the geometrical optics regime $\omega\gg\nu$ so that $s\simeq\omega/e$), we can extend the upper limit to infinity and evaluate the integral by analytic continuation of the standard identity \cite{Camblong:2020pme}
\begin{equation}
    \int_{0}^{\infty} dx\,x^{2i\nu} e^{ix}
    = -\frac{\pi e^{-\pi\nu}}{\sinh(2\pi\nu)\,\Gamma(-2i\nu)},
\end{equation}
with a correspondence of exponents $2i\nu\to -i\sigma$ and argument rescaling $x\to sx$. Following the same steps as in the Schwarzschild analysis and in related CQM treatments, one finds
\begin{equation}
    P_{\rm exc}
    \approx
    \frac{2\pi \gc^{2}\sigma}{e^{2} s^{2}}\,
    \frac{1}{e^{2\pi\sigma} - 1}.
    \label{eq:pex_thermal}
\end{equation}
This is a central result: the excitation probability contains a Planck factor $(e^{2\pi\sigma}-1)^{-1}$, whose argument is controlled by the near-horizon geometry through $F'_g$.

For an atom released from infinity, $e=1$. In the geometrical optics regime $\omega\gg\nu$, we can approximate $s\simeq\omega/e\simeq\omega$, which yields
\begin{equation}
    P_{\rm exc}
    \approx
    \frac{2\pi \gc^{2}\sigma}{\omega^{2}}\,
    \frac{1}{e^{2\pi\sigma} - 1},
    \qquad
    \sigma = \frac{2\nu}{F'_g}.
\end{equation}
It is convenient to rewrite the result in terms of the surface gravity $\kappa = F'_g/2$ and the Bardeen Hawking temperature $T_{H}^{(\mathrm{B})}=\kappa/(2\pi)$. Since
\begin{equation}
    2\pi\sigma
    = \frac{2\pi\nu}{\kappa}
    = \frac{\nu}{T_{H}^{(\mathrm{B})}},
\end{equation}
we can express the excitation probability as
\begin{equation}
    P_{\rm exc}
    \approx
    \frac{2\pi \gc^{2}}{\omega^{2}}
    \frac{\nu}{\kappa}\,
    \frac{1}{\exp\!\left(\dfrac{\nu}{T_{H}^{(\mathrm{B})}}\right) - 1}.
    \label{eq:pf}
\end{equation}
Equation~(\ref{eq:pf}) exhibits a thermal (Planckian) spectrum with temperature $T_{H}^{(\mathrm{B})}$ determined by the Bardeen surface gravity. The prefactor $\nu/\kappa$ encodes the density of modes in the near-horizon region and the dependence on the geometry, while the explicit $1/\omega^{2}$ factor reflects the suppression of excitation for large atomic gaps.

In the extremal limit $r_g^{2}\to 2g^{2}$, one has $\kappa\to 0$ and hence $T_{H}^{(\mathrm{B})}\to 0$, so that the excitation probability is strongly suppressed and the acceleration radiation effectively shuts off as the black hole approaches a cold remnant. Thus, the HBAR signal provides a quantum-optical diagnostic of the regular core parameter and of the approach to extremality in Bardeen-like geometries.

\subsection{Numerical behavior of the excitation probability}

\begin{figure*}[t]
\includegraphics[width=0.4\textwidth]{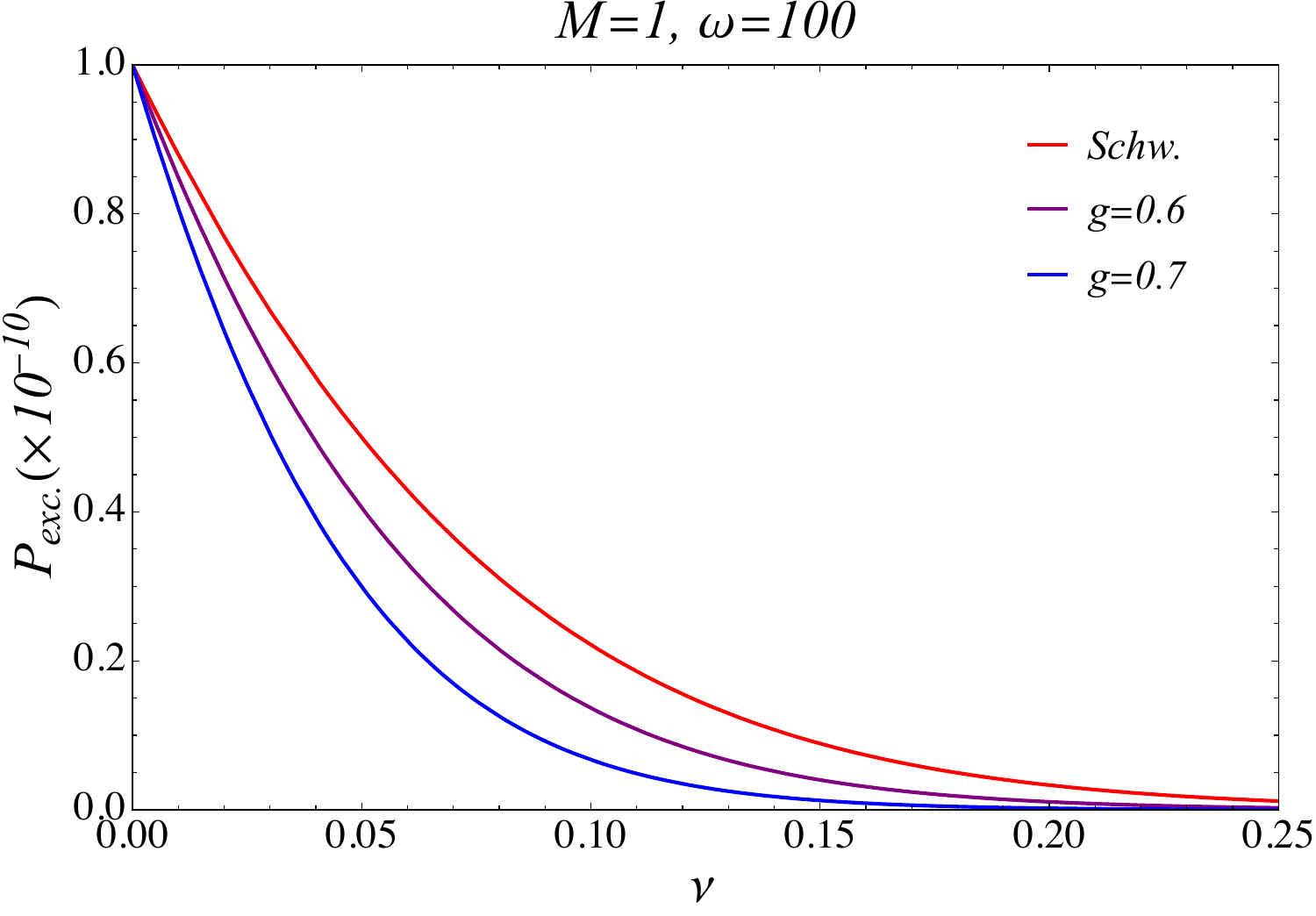}
\includegraphics[width=0.4\textwidth]{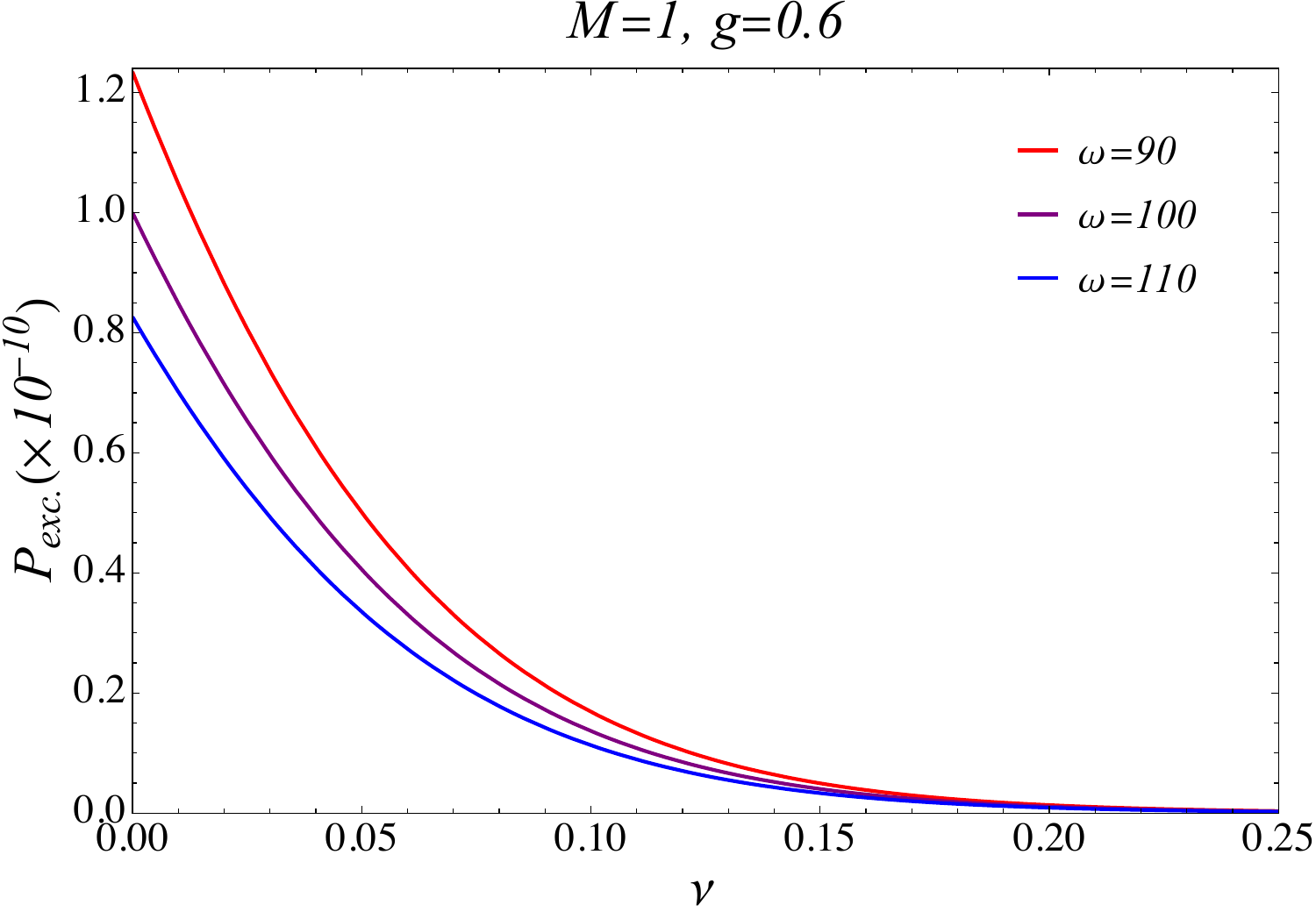}
\includegraphics[width=0.4\textwidth]{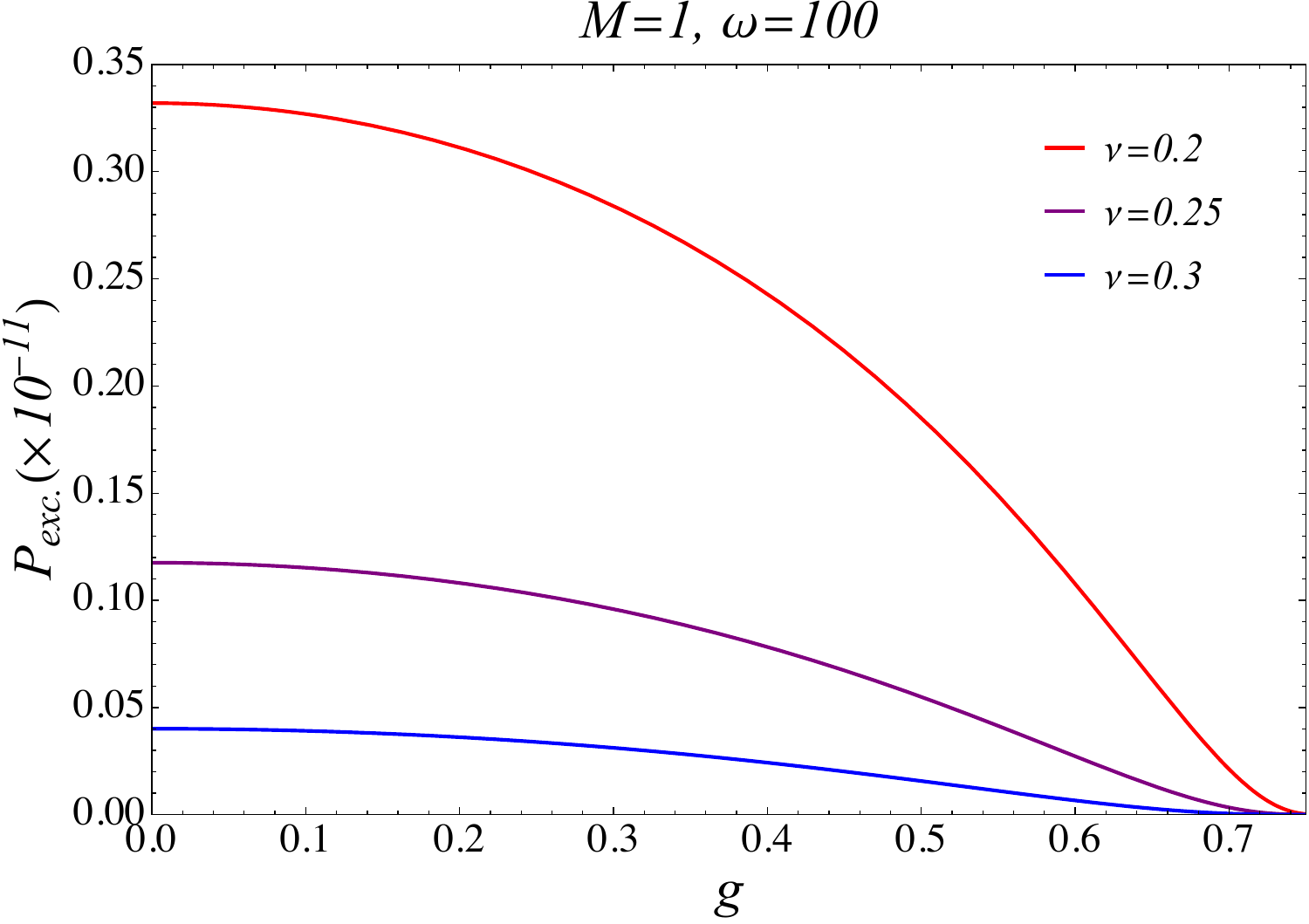}
\includegraphics[width=0.4\textwidth]{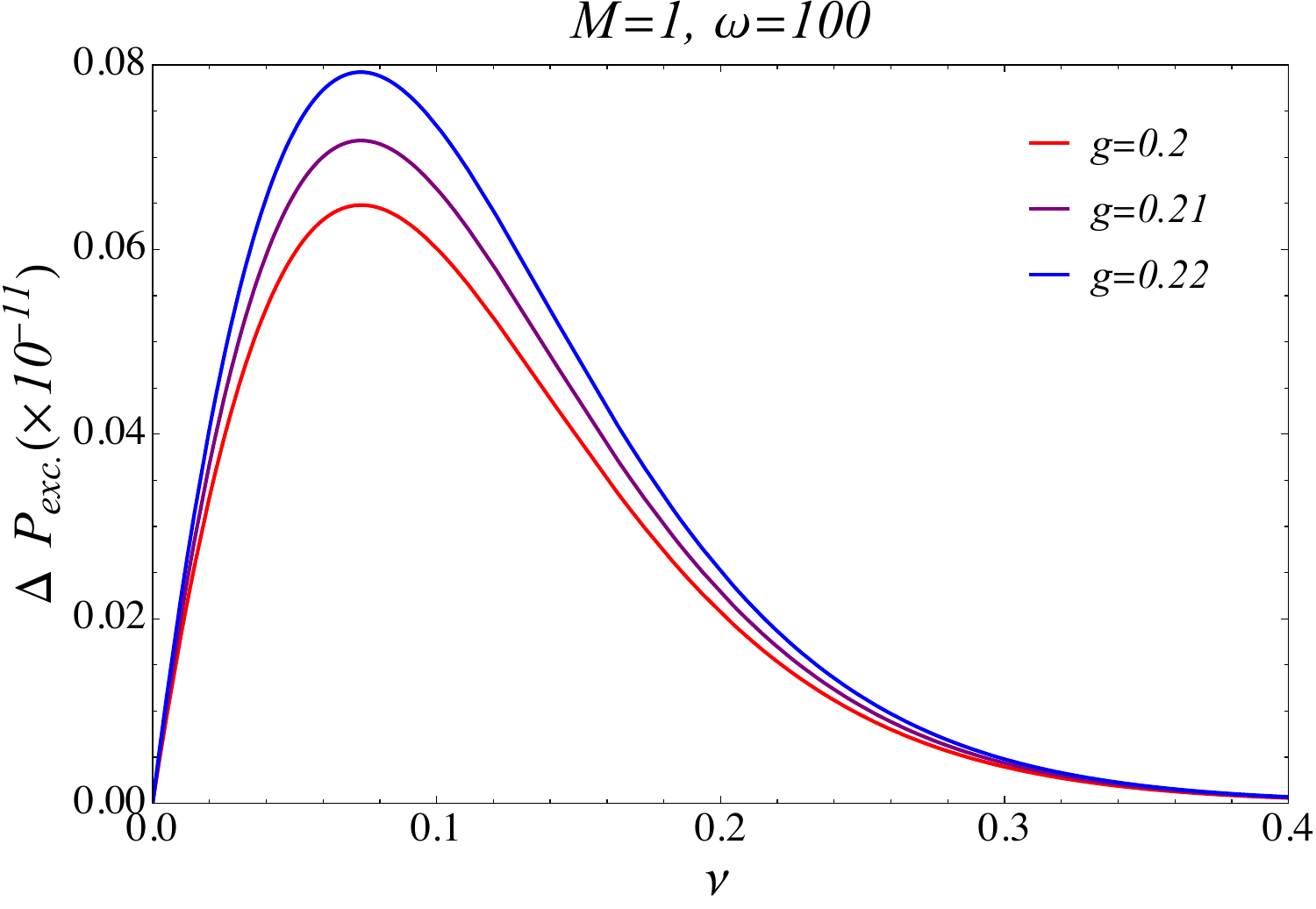}
\caption{The dependence of the radiation intensity $P_{exc}$ on the mode frequencies $\nu$ for different values of the Bardeen parameter $g$ (top left panel) and for different values of the atomic frequency $\omega$ (upper right panel). The radiation intensity $P_{exc}$ as a function of the Bardeen parameter  $g$ for different values of the $\nu$ (bottom left panel) and  
$\Delta P=P(0)-P(g)$ as a function $\nu$ (bottom right panel). Here we take $g_c=10^{-3}$.} \label{fig.P}
\end{figure*}

\begin{figure*}[t]
\includegraphics[width=0.6\textwidth]{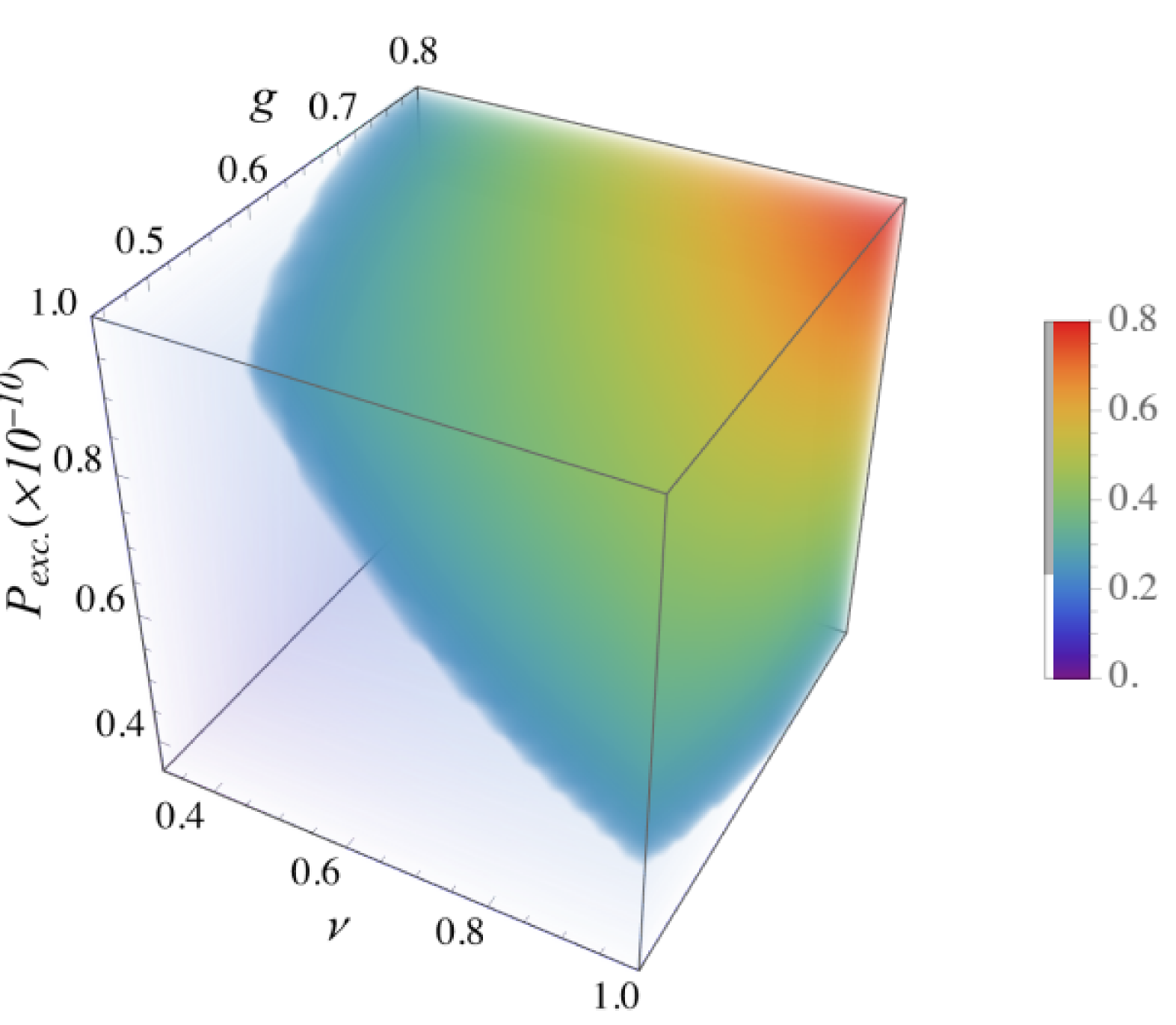}
\caption{
Radiation intensity $P_{\rm exc}$ plotted against mode frequencies $\nu$ and Bardeen parameter $g$ in 3 dimensions. \label{fig.3D P}}
\end{figure*}

Figure~2 summarizes how the excitation probability responds to variations of the
Bardeen parameter $g$, the mode frequency $\nu$, and the atomic gap $\omega$ in
the geometric-optics regime $\omega\gg\nu$ (here with $M=1$ and a small coupling
$\gc=10^{-3}$).  In frequency space, Eq.~\eqref{eq:pf} implies that
$P_{\rm exc}(\nu)$ approaches a constant as $\nu\to 0$ and is exponentially
suppressed for $\nu\gg T_H^{(\mathrm{B})}(g)$; the characteristic rollover scale
is set by the Bardeen Hawking temperature.  As $g$ increases, the surface gravity
$\kappa(g)$ (hence $T_H^{(\mathrm{B})}(g)$) decreases, suppressing the overall
amplitude and shifting the rollover toward smaller $\nu$, i.e.\ the regular core
``cools'' and dims the HBAR signal.

The upper-right panel, varying the atomic transition frequency (e.g.\ $\omega=90,100,110$ at fixed $g$)
does not change the thermal factor in $\nu$, but reduces the overall magnitude
through the explicit $\omega^{-2}$ prefactor in Eq.~\eqref{eq:pf}.  The dependence
on $g$ at fixed $(\nu,\omega)$ is monotonic: as $g$ approaches its extremal value
($\kappa\to 0^+$), the excitation probability is strongly quenched, consistent
with the vanishing Hawking temperature in that limit.


The lower-left panel isolates the dependence on the regular core parameter by plotting \(P_{\mathrm{exc}}\) as a function of \(g\) at fixed \(\omega = 100\) for several representative mode frequencies \(\nu\). For all \(\nu\) shown, the excitation probability decreases monotonically as \(g\) approaches its extremal value (where \(r_g^2 \to 2 g^2\) and \(\kappa(g)\to 0\)). This makes explicit that the Bardeen parameter directly controls the strength of acceleration radiation: as the horizon is softened and the black hole tends toward a cold remnant, the HBAR process is efficiently quenched.

Finally, the lower-right panel displays the difference
\begin{equation}
\Delta P_{\mathrm{exc}}(\nu;g) \equiv 
P_{\mathrm{exc}}^{\mathrm{Schw}}(\nu) - P_{\mathrm{exc}}^{\mathrm{Bard}}(\nu;g) \, ,
\end{equation}
for several nearby values of \(g\), again at fixed \(\omega = 100\). This quantity measures how much the regular geometry suppresses the excitation probability relative to the Schwarzschild case. The curves peak at low \(\nu\), where the thermal occupation is largest, and decay rapidly at higher frequencies. As \(g\) increases, the magnitude of \(\Delta P_{\mathrm{exc}}\) grows, signaling that the Bardeen deformation becomes increasingly effective at dimming the HBAR spectrum. Taken together, the four panels of Fig.~2 give a detailed quantitative view of how the Planckian spectrum of Eq.~(43) is reshaped by the regular core parameter and by the atomic gap.

\vspace{0.3cm}

Figure~3 offers a complementary two- and three-dimensional view of the excitation probability and highlights the regions of parameter space where HBAR is most efficient. When \(P_{\mathrm{exc}}\) is plotted as a function of \(\nu\) and \(g\) (for fixed \(M\) and \(\omega\)), the largest values are concentrated in the corner corresponding to small mode frequency and small Bardeen parameter. Along the \(\nu\)-direction, slices at fixed \(g\) reproduce the Planckian spectrum discussed in Fig.~2, with a maximum at \(\nu \sim T_{H}^{(B)}(g)\) and an exponentially damped high-frequency tail. Along the \(g\)-direction, slices at fixed \(\nu\) show a rapid suppression of the signal as the core scale grows and the surface gravity \(\kappa(g)\) decreases. Near the extremal regime the entire surface collapses toward \(P_{\mathrm{exc}} \approx 0\), illustrating in a single plot how the approach to a cold regular remnant switches off horizon-brightened acceleration radiation.

When the same quantity is instead viewed as a function \(P_{\mathrm{exc}}(\nu,\omega)\) at fixed \(g\), the strong interplay between the field and atomic frequency scales becomes apparent. The largest excitation probabilities are concentrated where both \(\nu\) and \(\omega\) are small, i.e., low-frequency modes interacting with low-gap atoms. Moving away from this region, the probability decays rapidly with increasing \(\nu\), due to the thermal factor \(\exp[-\nu/T_{H}^{(B)}(g)]\), and with increasing \(\omega\), due to the algebraic \(\omega^{-2}\) prefactor. This representation makes it clear that the HBAR process is dominated by soft quanta and ``gentle'' detectors, while high-frequency modes or atoms with large level spacing are strongly disfavored.

Overall, Fig.~3 visualizes in a compact way the main analytic conclusions of Sec.~V: the excitation probability is thermal in the mode frequency with an effective temperature set by the Bardeen surface gravity, and its magnitude is controlled both by the regular core parameter \(g\) and by the atomic microphysics encoded in \(\omega\).

\section{HBAR entropy and area law in the Bardeen geometry}
\label{sec:HBARentropy}

In the previous section we showed that the excitation probability of a freely falling two-level atom in the Bardeen spacetime is Planckian in the mode frequency~$\nu$, with an effective temperature equal to the Bardeen Hawking temperature $T_H^{(\mathrm{B})}$,
\begin{eqnarray}
 P_{\rm exc}(\nu,\omega;g)
 \simeq
 \frac{2\pi \gc^{2}}{\omega^{2}}\,
 \frac{\nu}{\kappa(g)}\,
 \frac{1}{\exp\!\left(\dfrac{\nu}{T_{H}^{(\mathrm{B})}(g)}\right)-1},
 \\
 T_{H}^{(\mathrm{B})}(g)=\frac{\kappa(g)}{2\pi},
 \label{eq:Pexc_HB}
\end{eqnarray}
where $\omega$ is the atomic gap and $\kappa(g)=\tfrac12 F'_g$ is the surface gravity of the Bardeen black hole [see Eq.~\eqref{eq:kappa_T}].  This Planck factor implies that a single cavity mode interacting with a flux of such atoms will thermalize at the Bardeen Hawking temperature, which in turn leads to a natural notion of \emph{horizon-brightened acceleration radiation} (HBAR) entropy.

In this section we adapt the quantum-optics derivation of HBAR entropy \`a la Scully et al.\ and the conformal-quantum-mechanics (CQM) framework of Refs.~\cite{Scully:2017utk,Azizi:2021qcu,Azizi:2021yto,Ordonez:2025sqp} to the Bardeen geometry.  The result is an HBAR entropy flux that satisfies the same area law as in the Schwarzschild case, but with all geometric dependence encoded in the Bardeen surface gravity and horizon radius.  We emphasize the role of the Bardeen parameter $g$ and show that both the HBAR energy and entropy fluxes are quenched in the extremal Bardeen limit.

\subsection{Single-mode master equation and detailed balance}

We consider one cavity mode of Killing frequency $\nu>0$, with annihilation and creation operators $a$ and $a^{\dagger}$.  Two-level atoms with gap $\omega$ are injected into the cavity at an average flux $\mathcal{J}$ (number of atoms per unit Killing time).  Each atom follows a radial geodesic and interacts with the scalar field along its worldline via the interaction Hamiltonian~\eqref{eq:v}.  For a given mode, the probability that one atom undergoes the transition
\(
|b,0_\nu\rangle\to|a,1_\nu\rangle
\)
is $P_{\rm exc}(\nu,\omega;g)$ in Eq.~\eqref{eq:Pexc_HB}.  The corresponding probability for the reverse process,
\(
|a,1_\nu\rangle\to|b,0_\nu\rangle
\),
is $P_{\rm abs}(\nu,\omega;g)$.

Following the standard HBAR laser-like analysis, the mode density matrix $\rho(t)$ in the interaction picture obeys a Lindblad master equation
\begin{equation}
 \dot{\rho}
 =
 -i\,[H_{\rm eff},\rho]
 + \Gamma_{\rm em}\,\mathcal{D}[a^{\dagger}]\rho
 + \Gamma_{\rm abs}\,\mathcal{D}[a]\rho ,
 \label{eq:master}
\end{equation}
where $H_{\rm eff}=\nu\, a^{\dagger}a$ is the effective Hamiltonian, the dissipator is
\begin{equation}
 \mathcal{D}[O]\rho
 = O\,\rho\,O^{\dagger}
 - \frac{1}{2}\left\{O^{\dagger}O,\rho\right\},
\end{equation}
and the emission and absorption rates are
\begin{equation}
 \Gamma_{\rm em}(\nu;g) = \mathcal{J}\,P_{\rm exc}(\nu,\omega;g),
 \,\,
 \Gamma_{\rm abs}(\nu;g) = \mathcal{J}\,P_{\rm abs}(\nu,\omega;g).
 \label{eq:rates_def}
\end{equation}
Because the excitation probability already includes the Planck factor at $T_{H}^{(\mathrm{B})}$, detailed balance fixes the ratio of the rates to be
\begin{equation}
 \frac{\Gamma_{\rm em}}{\Gamma_{\rm abs}}
 = e^{-\beta_{H}^{(\mathrm{B})}\nu},
 \qquad
 \beta_{H}^{(\mathrm{B})}
 = \frac{1}{T_{H}^{(\mathrm{B})}(g)}
 = \frac{2\pi}{\kappa(g)}.
 \label{eq:detailed_balance}
\end{equation}
Equivalently, $\Gamma_{\rm abs}=\Gamma_{\rm em}\,e^{\beta_{H}^{(\mathrm{B})}\nu}$.

Writing the master equation in the Fock basis, the diagonal elements $\rho_{n,n}(t)$ obey
\begin{eqnarray}
\dot{\rho}_{n,n}
&=& \Gamma_{\rm em}\Bigl[n\,\rho_{n-1,n-1}-(n+1)\rho_{n,n}\Bigr]
\\\nonumber
&+& \Gamma_{\rm abs}\Bigl[(n+1)\rho_{n+1,n+1}-n\,\rho_{n,n}\Bigr].
\label{eq:rate_rho}
\end{eqnarray}

Imposing stationarity, $\dot{\rho}_{n,n}=0$, yields
\begin{equation}
\frac{\rho^{(\mathrm{st})}_{n,n}}{\rho^{(\mathrm{st})}_{n-1,n-1}}
=\frac{\Gamma_{\rm em}}{\Gamma_{\rm abs}}
\equiv q \, ,
\qquad 0<q<1,
\end{equation}
so the normalized solution is
\begin{equation}
\rho^{(\mathrm{st})}_{n,n}=(1-q)\,q^{n}.
\end{equation}
Using detailed balance given in Eq. \eqref{eq:detailed_balance}, we obtain the Bose-Einstein steady state
\begin{equation}
\rho^{(\mathrm{st})}_{n,n}
=\bigl(1-e^{-\beta_{H}^{(\mathrm{B})}\nu}\bigr)\,e^{-n\beta_{H}^{(\mathrm{B})}\nu},
\,\,
\bar n_\nu
=\sum_n n\,\rho^{(\mathrm{st})}_{n,n}
=\frac{1}{e^{\beta_{H}^{(\mathrm{B})}\nu}-1}.
\label{eq:rho_thermal}
\end{equation}

Thus, for each Bardeen background characterized by $(m,g)$, the cavity mode thermalizes to the Bardeen Hawking temperature $T_H^{(\mathrm{B})}(g)$, in complete analogy with the Schwarzschild case.

\subsection{HBAR energy and entropy flux}

The HBAR energy stored in the radiation field is
\begin{equation}
 E_{P}(t)
 = \sum_{\nu} \nu\,\bar{n}_{\nu}(t),
 \qquad
 \dot{E}_{P}
 = \sum_{\nu} \nu\,\dot{\bar{n}}_{\nu},
 \label{eq:EP_def}
\end{equation}
where the sum over $\nu$ can be replaced by an integral in the continuum limit.

From the master equation~\eqref{eq:master} one finds for the mode occupation number
\begin{equation}
\dot{\bar n}_{\nu}
=\Gamma_{\rm em}\,(\bar n_{\nu}+1)-\Gamma_{\rm abs}\,\bar n_{\nu}
=\Gamma_{\rm em}-(\Gamma_{\rm abs}-\Gamma_{\rm em})\,\bar n_{\nu}.
\end{equation}
In the early-time (or dilute) regime $\bar n_\nu\ll 1$, this reduces to
\begin{equation}
\dot{\bar{n}}_{\nu}\simeq \Gamma_{\rm em}(\nu;g)
=\mathcal{J}\,P_{\rm exc}(\nu,\omega;g),
\label{eq:nbar_dot_approx}
\end{equation}
which we use below to estimate the initial HBAR energy/entropy production rates.

Combining Eqs.~\eqref{eq:Pexc_HB}, \eqref{eq:EP_def}, and~\eqref{eq:nbar_dot_approx} yields the HBAR energy flux in the Bardeen geometry,
\begin{equation}
 \dot{E}_{P}(g)
 \simeq
 \frac{2\pi \gc^{2}\mathcal{J}}{\omega^{2}}
 \sum_{\nu}
 \frac{\nu^{2}}{\kappa(g)}\,
 \frac{1}{\exp\!\left(\dfrac{\nu}{T_{H}^{(\mathrm{B})}(g)}\right)-1}.
 \label{eq:EP_flux}
\end{equation}
The dependence on the Bardeen parameter $g$ enters exclusively through the surface gravity $\kappa(g)$ and the Hawking temperature $T_{H}^{(\mathrm{B})}(g)$.

The von Neumann entropy of the radiation field is
\begin{equation}
 S_{\rho}(t)
 = -\sum_{n,\nu} \rho_{n,n}(t)\,\ln\rho_{n,n}(t),
 \label{eq:Srho_def}
\end{equation}
and its rate of change due to photon emission is
\begin{equation}
 \dot{S}_{\rho}
 = -\sum_{n,\nu} \dot{\rho}_{n,n}\,\ln \rho_{n,n}^{(\mathrm{st})}.
 \label{eq:Srho_dot_def}
\end{equation}
Using the master equation together with the steady-state distribution~\eqref{eq:rho_thermal}, one finds, exactly as in the Schwarzschild case,
\begin{equation}
 \dot{S}_{\rho}
 = \beta_{H}^{(\mathrm{B})}\,\dot{E}_{P},
 \label{eq:HBAR_first_law}
\end{equation}
which is the HBAR analogue of the Clausius relation and exhibits the direct proportionality between HBAR entropy and energy flux.

Equation~\eqref{eq:HBAR_first_law} shows that the HBAR entropy production rate is controlled by the same inverse temperature $\beta_H^{(\mathrm{B})}$ that characterizes the Planckian spectrum in Eq.~\eqref{eq:Pexc_HB}, with all geometric information encoded in the Bardeen surface gravity.  Explicitly,
\begin{equation}
 \beta_{H}^{(\mathrm{B})}(g)
 = \frac{1}{T_{H}^{(\mathrm{B})}(g)}
 = \frac{2\pi r_g\bigl(r_g^{2}+g^{2}\bigr)}{\tfrac12 r_g^{2}-g^{2}},
 \label{eq:beta_B}
\end{equation}
where $r_g=r_g(m,g)$ is the outer horizon radius determined by Eq.~\eqref{eq:rg_def}.

\subsection{Wien’s displacement law in the Bardeen background}
\label{sec:Wien_Bardeen}

The Planckian spectrum~\eqref{eq:Pexc_HB} for the HBAR excitation probability implies the existence of a Wien-type displacement law in the Bardeen geometry. As in ordinary blackbody radiation, there is a characteristic wavelength at which the spectrum attains its maximum, and the product of this wavelength with the effective temperature is a universal constant determined solely by the functional form of the spectrum \cite{ScullyHBAR,Das:2023rwg,Jana:2024fhx}. In the present context, the relevant temperature is the Bardeen Hawking temperature $T_{H}^{(\mathrm{B})}(g)$, which encodes all dependence on the regular core parameter $g$ through the surface gravity. In this subsection we make this statement precise by recasting the excitation probability in wavelength space, deriving the corresponding displacement law, and analyzing how the spectral peak responds to variations of $g$.

\subsubsection{Spectral profile in wavelength space}

We regard the excitation probability $P_{\rm exc}(\nu,\omega;g)$ as an effective spectral weight for the HBAR process in mode frequency $\nu$. Up to a frequency-independent prefactor (containing $\gc$, $\omega$ and the injection flux $\mathcal{J}$), the relevant $\nu$-dependence is
\begin{equation}
 \mathcal{P}_{\nu}(\nu;g)
 \propto
 \frac{\nu}{\exp\!\left(\dfrac{\nu}{T_{H}^{(\mathrm{B})}(g)}\right)-1},
 \label{eq:Pnu_spectrum}
\end{equation}
which has the same structure as the standard Planck spectrum in frequency space.

To obtain the corresponding profile in terms of the wavelength $\lambda$, we use $\lambda = 1/\nu$ in natural units. The spectral weight per unit wavelength, $\mathcal{P}_{\lambda}(\lambda;g)$, is defined by
\begin{eqnarray}
 \mathcal{P}_{\lambda}(\lambda;g)\,d\lambda
 &=& \mathcal{P}_{\nu}(\nu;g)\,d\nu,
 \qquad
 \nu = \frac{1}{\lambda},
\\
 d\nu &=&-\frac{1}{\lambda^{2}}\,d\lambda.
\end{eqnarray}
Dropping overall signs and normalization factors (which do not influence the position of the maximum), we arrive at
\begin{equation}
 \mathcal{P}_{\lambda}(\lambda;g)
 \propto
 \frac{1}{\lambda^{3}}\,
 \frac{1}{\exp\!\left(\dfrac{1}{\lambda T_{H}^{(\mathrm{B})}(g)}\right)-1}.
 \label{eq:Plambda_spectrum}
\end{equation}
Equation~\eqref{eq:Plambda_spectrum} is the analogue, for the HBAR excitation probability, of the familiar Planck law written in wavelength space. All geometric information associated with the Bardeen metric is encoded in the single parameter $T_{H}^{(\mathrm{B})}(g)$.

\subsubsection{Displacement law and critical wavelength}

The peak of the spectrum~\eqref{eq:Plambda_spectrum} is determined by the extremum condition 
\begin{equation}
 \frac{d\mathcal{P}_{\lambda}(\lambda;g)}{d\lambda}\bigg|_{\lambda=\lambda_{\rm crit}} = 0.
 \label{eq:extremum_condition}
\end{equation}
To make the temperature dependence manifest, we introduce the dimensionless variable
\begin{equation}
 x
 = \frac{1}{\lambda T_{H}^{(\mathrm{B})}(g)},
 \qquad
 \lambda
 = \frac{1}{x\,T_{H}^{(\mathrm{B})}(g)}.
 \label{eq:x_def}
\end{equation}
In terms of $x$, the spectrum becomes
\begin{equation}
 \mathcal{P}_{\lambda}(x)
 \propto
 \frac{x^{3}}{\exp(x)-1},
\end{equation}
and the condition~\eqref{eq:extremum_condition} translates into
\begin{equation}
 \frac{d}{dx}\left[
   \frac{x^{3}}{e^{x}-1}
 \right]_{x=x_{\rm crit}} = 0.
\end{equation}
Evaluating the derivative yields the transcendental equation
\begin{equation}
 1 - e^{-x_{\rm crit}} = \frac{x_{\rm crit}}{3},
 \label{eq:Wien_equation}
\end{equation}
whose unique positive solution is
\begin{equation}
 x_{\rm crit} \simeq 2.82144.
\end{equation}
Substituting back into Eq.~\eqref{eq:x_def}, we find the Wien displacement law for the Bardeen HBAR spectrum,

\begin{equation}
\lambda_{\rm crit}^{(\mathrm{B})} T_H^{(\mathrm{B})}(g)
 = \frac{1}{x_{\rm crit}}
 \simeq 0.3544,
 \label{eq:Wien_Bardeen}
\end{equation}

in natural units. Restoring $\hbar$ and $k_{B}$ yields
\begin{equation}
\lambda_{\rm crit}^{(\mathrm{B})}
 \simeq
 \frac{2\pi}{x_{\rm crit}}
 \frac{\hbar}{k_B\,T_H^{(\mathrm{B})}(g)}
 \approx 2.23\,\frac{\hbar}{k_B\,T_H^{(\mathrm{B})}(g)}.
 \label{eq:Wien_Bardeen_units}
\end{equation}

The numerical constant is identical to the one appearing in the Schwarzschild and quantum-corrected cases, reflecting the fact that it is determined solely by the shape of the Planck factor. The Bardeen geometry affects only the value of $T_{H}^{(\mathrm{B})}(g)$, and hence shifts the overall scale of the spectrum without altering the universal displacement constant.

\begin{figure*}[ht!]
\includegraphics[width=0.38\textwidth]{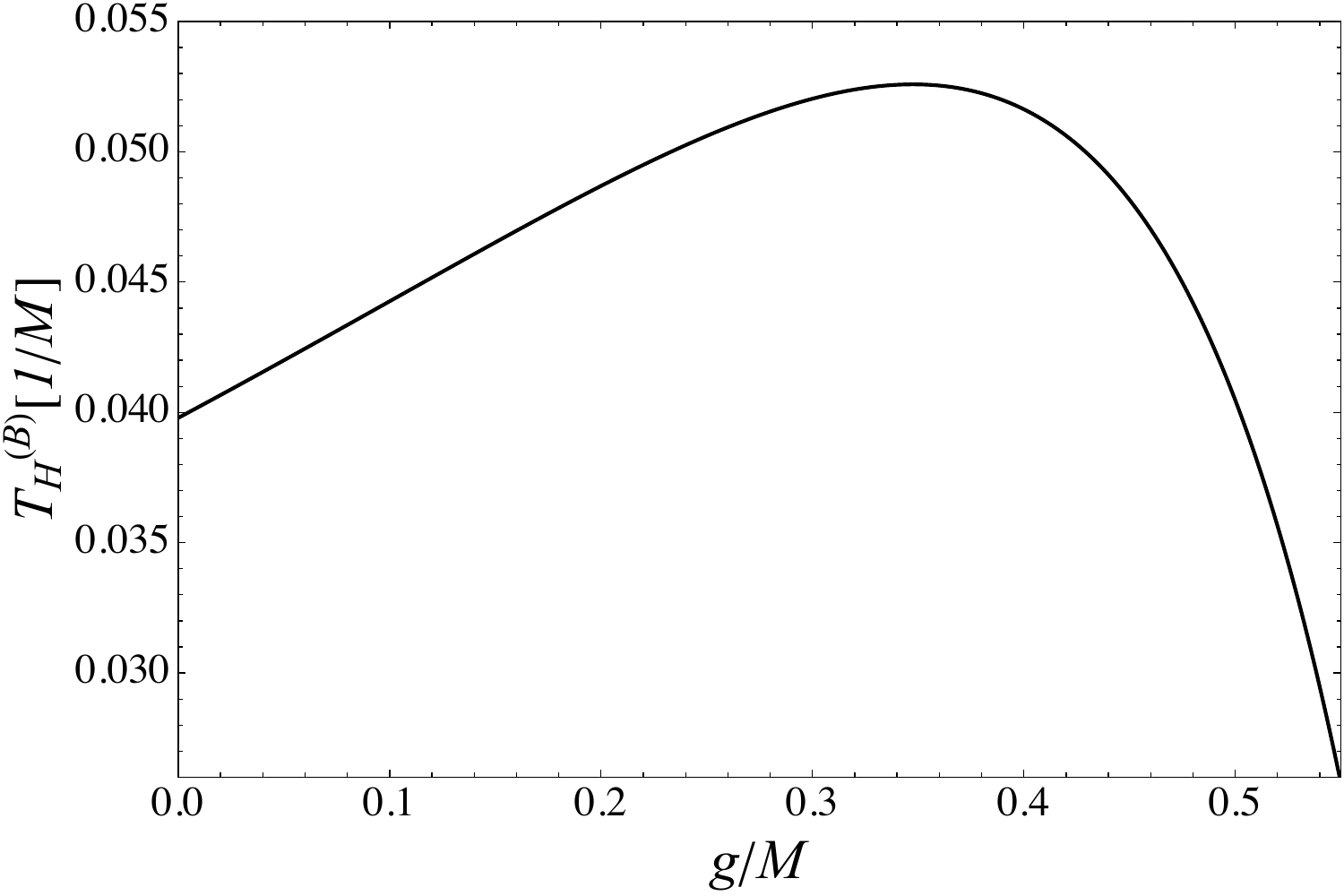}
\includegraphics[width=0.4\textwidth]{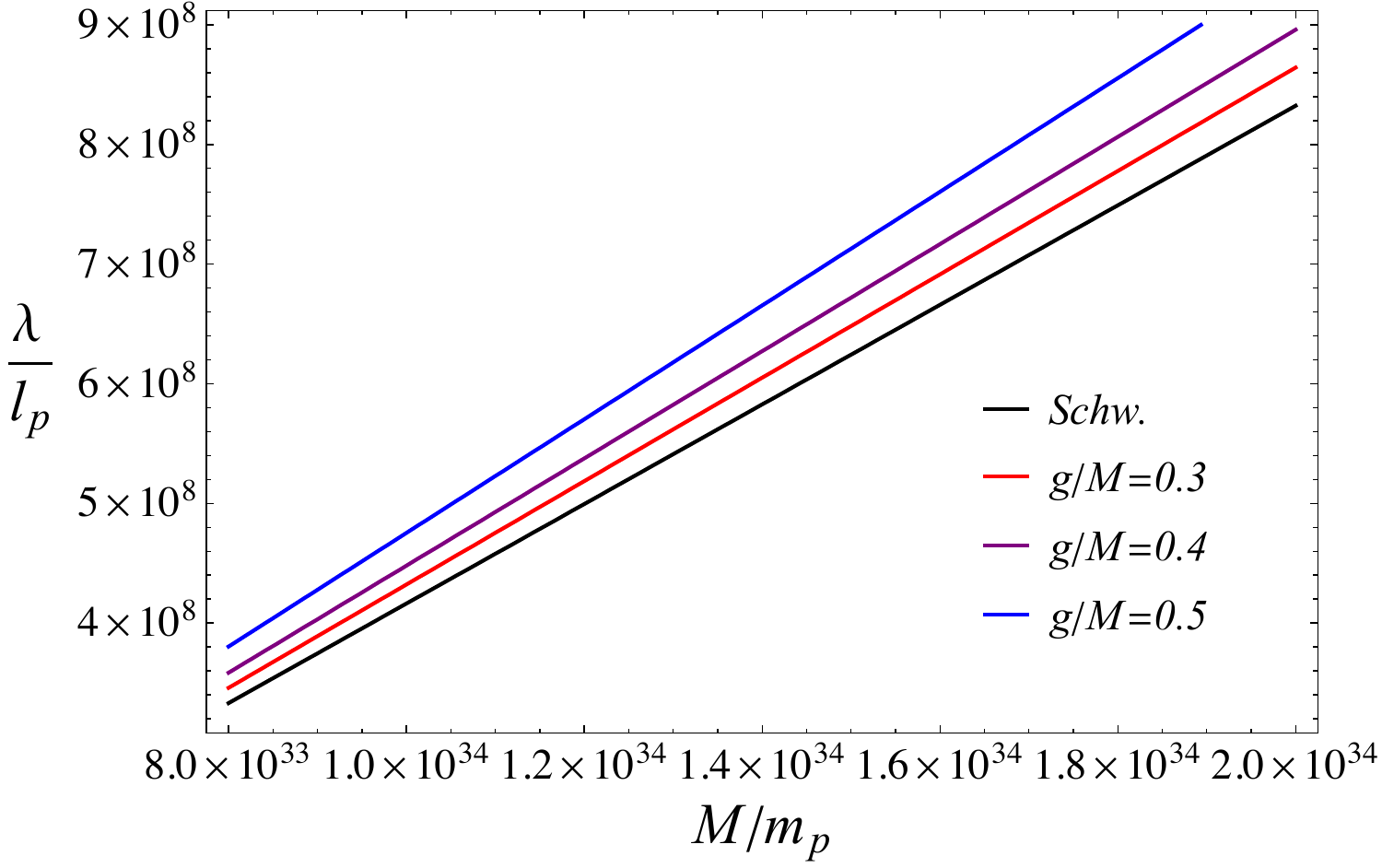}
\caption{
Dependence of the Hawking temperature $T_H^{(B)}$ on the Bardeen parameter $g$ (left plot) and dependence  of the wavelength ($\lambda/l_p$) on the  mass ($M/m_p$) of the black hole (right plot) from the Wien’s displacement law . Here $l_p$ and $m_p$ indicate the Planck length and Planck mass respectively.} \label{fig.Wien}
\end{figure*}

Figure~\ref{fig.Wien} illustrates how the Bardeen regularization parameter $g$
modifies the thermodynamic properties of the black hole and, in turn, the
characteristic wavelength associated with the thermal scale set by
$T_H^{(\mathrm{B})}$. In the left panel we plot $T_H^{(\mathrm{B})}$ versus $g/M$.
As $g/M$ increases toward the extremal value, the surface gravity decreases and
$T_H^{(\mathrm{B})}\to 0$, signaling the approach to a cold remnant. Consequently,
the corresponding characteristic (peak) wavelength inferred from Wien’s law
shifts to larger values, reflecting a progressively softer spectrum near
extremality.

The right panel shows the corresponding dependence of the characteristic (peak)
wavelength of the Hawking spectrum on the black-hole mass \(M/m_p\), obtained from
Wien's displacement law. The approximately linear growth with \(M\) is the expected
trend for black-hole radiation: heavier black holes are colder and therefore emit
at longer wavelengths. The key new feature is the systematic upward shift of the
curves as \(g/M\) increases: for a fixed mass, Bardeen black holes have a longer
peak wavelength than the Schwarzschild case, i.e.\ their spectrum is shifted toward
softer (redder) quanta. This is a direct imprint of the reduced effective horizon
temperature produced by the regular core at sufficiently large \(g/M\). In this
sense, the parameter \(g\) provides a clean thermodynamic handle on the spectrum:
increasing \(g/M\) progressively reddens the Hawking emission and pushes the system
toward a regime where the characteristic wavelength becomes very large, consistent
with suppressed radiation near extremality. These trends are also consistent with
the behavior of excitation probabilities discussed earlier: a cooler horizon implies
a weaker population of high-frequency modes and hence a reduced efficiency for
horizon-driven radiative processes such as HBAR.

\subsubsection{Dependence on the regular core parameter}

Equation~\eqref{eq:Wien_Bardeen_units} makes the role of the regular core parameter $g$ completely transparent. For fixed ADM mass $m$, increasing $g$ lowers the surface gravity $\kappa(g)$ and therefore the Bardeen Hawking temperature $T_{H}^{(\mathrm{B})}(g)$. Consequently, the critical wavelength behaves as
\begin{eqnarray}
 g_{1} < g_{2}
 \quad\Rightarrow\quad
 T_{H}^{(\mathrm{B})}(g_{1}) > T_{H}^{(\mathrm{B})}(g_{2})
 \\\quad\Rightarrow\quad 
 \lambda_{\rm crit}^{(\mathrm{B})}(g_{1})
 < \lambda_{\rm crit}^{(\mathrm{B})}(g_{2}).
\end{eqnarray}
In other words, as the regular core grows, the HBAR spectrum becomes colder and redder: the peak shifts toward longer wavelengths while the overall excitation probability is suppressed.

In the extremal limit $r_g^{2}\to 2g^{2}$, one has $\kappa(g)\to 0$ and $T_{H}^{(\mathrm{B})}(g)\to 0$, so that
\begin{equation}
 \lambda_{\rm crit}^{(\mathrm{B})}(g)
 \to \infty,
 \qquad
 T_{H}^{(\mathrm{B})}(g)\to 0.
\end{equation}
The displacement law thus encapsulates, in a particularly simple and quantitative way, how the Bardeen parameter controls the spectral content of acceleration radiation: regularization of the core softens the horizon, quenches the HBAR signal, and drives the characteristic wavelength to ever larger values as the black hole approaches a cold remnant.

From the broader HBAR perspective, the result~\eqref{eq:Wien_Bardeen_units} shows that the structure of Wien’s displacement law is insensitive to singularity resolution at the metric level; its only imprint is the $g$-dependence of the Hawking temperature. This is fully consistent with the entropy analysis of Sec.~\ref{sec:HBARentropy}, where the Bardeen geometry preserves the area/4 law without logarithmic corrections, and reinforces the conclusion that the dominant near-horizon physics probed by freely falling detectors is governed by the surface gravity and the associated conformal quantum mechanics.

\section{Conclusions}
\label{conclusion}

In this work we have calculated the HBAR of Bardeen regular black hole spacetimes. Starting from the Bardeen line element, we characterized the horizon structure, surface gravity and Hawking temperature in terms of the core parameter that controls the size of the de Sitter-like regular region. On this background we analyzed the radial geodesics of freely falling atoms and derived near-horizon expansions of proper and coordinate time, identifying the leading behavior that governs the detector response in the vicinity of the horizon.

We then studied a minimally coupled massless scalar field and showed that, in the near-horizon region and for s-wave modes, the radial Klein-Gordon equation once again reduces to an effective conformal quantum mechanics problem with an inverse-square potential. The effective coupling is fixed entirely by the Bardeen surface gravity, demonstrating that the conformal structure responsible for HBAR thermality is robust under singularity resolution: the replacement of the central curvature singularity by a regular de Sitter-like core does not spoil the universal near-horizon CQM dynamics, but instead feeds in only through the modified surface gravity.

Coupling this field to a freely falling two-level atom in a Boulware-like vacuum with a stretched-horizon mirror, we computed the excitation probability using time-dependent perturbation theory. The resulting spectrum is thermal in the mode frequency, with a temperature equal to the Hawking temperature of the Bardeen black hole. The regular core parameter controls the overall strength and shape of the spectrum: as it increases, the surface gravity and Hawking temperature are reduced, the excitation probability is strongly suppressed, and the HBAR signal fades as the geometry approaches the extremal, cold-remnant limit. Our numerical analysis confirms these trends and illustrates in detail how low-frequency modes and small atomic gaps dominate the HBAR response, while high-frequency modes and large gaps are exponentially or algebraically suppressed.

Beyond the single-atom response, we constructed a coarse-grained master equation for a single cavity mode interacting with a flux of infalling atoms in the Bardeen background. The emission and absorption rates satisfy detailed balance at the Bardeen Hawking temperature, and the mode relaxes to a thermal Bose-Einstein state. The associated HBAR energy and entropy fluxes obey a Clausius-type relation, with all geometric dependence entering through the Bardeen surface gravity and horizon radius. In this way, the HBAR entropy in regular Bardeen spacetimes retains the same area-law character as in the Schwarzschild case, but with a prefactor that is smoothly deformed by the regular core scale and quenched in the extremal limit.

Taken together, these results show that regular black holes provide a natural and fertile arena for the HBAR and CQM program. The Bardeen parameter plays the role of a tunable regularization scale that simultaneously controls the surface gravity, the strength of acceleration radiation and the structure of the associated entropy, turning HBAR into a quantum-optical probe of regular black hole interiors and remnant thermodynamics. Future work can extend this analysis to other regular metrics, to rotating and charged regular black holes, to different field content and detector couplings, and to multi-mode or non-Markovian regimes. On the experimental side, it will be interesting to explore whether analogue-gravity platforms and engineered quantum-optical setups can emulate the effective near-horizon dynamics found here and thereby shed light on the phenomenology of regular horizons and their possible quantum-gravity origin.

\acknowledgments
A. \"O. and R. P. would like to acknowledge networking support of the COST Action CA21106 - COSMIC WISPers in the Dark Universe: Theory, astrophysics and experiments (CosmicWISPers), the COST Action CA22113 - Fundamental challenges in theoretical physics (THEORY-CHALLENGES), the COST Action CA21136 - Addressing observational tensions in cosmology with systematics and fundamental physics (CosmoVerse), the COST Action CA23130 - Bridging high and low energies in search of quantum gravity (BridgeQG), and the COST Action CA23115 - Relativistic Quantum Information (RQI) funded by COST (European Cooperation in Science and Technology). A. \"O. would also like to acknowledge the funding support of SCOAP3, Switzerland and TUBITAK, Turkiye.

\bibliography{ref}
\end{document}